\documentclass[]{spie}  %>>> use for US letter paper

\newcommand{\arcsec}{\ensuremath{^{\prime\prime}}}

\usepackage{amsmath,amsfonts,amssymb}
\usepackage{graphicx}
\usepackage{subfigure}
\usepackage[colorlinks=true, allcolors=blue]{hyperref}
\usepackage{svg}
\usepackage{gensymb}
\usepackage{amssymb}
\usepackage{wasysym}
\usepackage{textcomp}
\usepackage{tcolorbox}
\usepackage{graphicx}	% Including figure files
\usepackage{amsmath}	% Advanced maths commands
\usepackage{subcaption}
\usepackage{hyperref}
\usepackage{float}
\usepackage{siunitx}
\usepackage{hyperref}
\usepackage{url}
\usepackage{subfigure}
\usepackage{caption}
\title{An Initial Assessment of the Hanle Echelle Spectrograph \\ for \\ Exoplanet Atmosphere Studies}

\author[a,b]{Manjunath Bestha}
\author[c]{Athira Unni}
\author[a]{T. Sivarani}
\author[d]{Shivangi Menon}
\author[a,b]{Parvathy M}
\author[a]{Arun Surya}
\author[a,e]{Pallavi Saraf}
\author[f]{Devika Divakar}
\author[a]{Lokesh Manickavasaham}

\affil[a]{Indian Institute of Astrophysics, Bangalore, India}
\affil[b]{University of Calcutta, India}
\affil[c]{University of California, Santa Cruz, USA}
\affil[d]{Indian Institute of Science Education and Research, Kolkata, India}
\affil[e]{Physical Research Laboratory, Ahmadabad, India}
\affil[f]{University of Texas, Austin, USA}

\authorinfo{Further author information(Send correspondence to Manjunath Bestha): \\ Manjunath Bestha: E-mail: bestha95@gmail.com\\  Sivarani Thirupathi: E-mail: sivarani@iiap.res.in}

\begin{document} 

\maketitle

\begin{abstract}

Transmission spectroscopy is an effective technique for probing exoplanetary atmospheres. While most observations have relied on space facilities such as HST and JWST, ground-based high-resolution transmission spectroscopy (HRTS) has also provided valuable insights by resolving individual atomic features. In this work, we present an initial performance assessment and feasibility test of the Hanle Echelle Spectrograph (HESP) on the 2 m Himalayan Chandra Telescope (HCT) for HRTS. As a benchmark, we observed the hot Jupiter HD 209458b during a single transit at a resolution of R = 30,000. We developed a Python-based, semi-automated data reduction and analysis pipeline that includes corrections for telluric contamination and stellar radial velocity shifts. The final achieved signal-to-noise ratio and spectral stability allow us to probe for features at the ~0.1\% level. This work establishes a methodology and demonstrates the operational capability of the HESP-HCT for obtaining high-resolution transmission spectra.

\end{abstract}


\section{Introduction}
\label{section1}

High-resolution transmission spectroscopy (HRTS) is a powerful tool for characterizing exoplanet atmospheres, allowing the detection of atomic and molecular species and the study of their dynamics through resolved line profiles. This technique allows for the study of physical processes such as wind patterns, thermal structures, and escape mechanisms, especially in close-in gas giants where stellar irradiation drives complex atmospheric behavior \cite{Wyttenbach_2015...577A..62W, Nugroho_2017AJ....154..221N, Yan_2018NatAs...2..714Y}.

While large-aperture facilities have traditionally dominated this technique, recent efforts have demonstrated that even modest-sized telescopes ranging from 1.5 m to 2.5 m in aperture equipped with high-resolution spectrographs can yield scientifically valuable results, particularly for bright targets and carefully optimized observing strategies \cite{Bellow_1.5m, Lowson_2023_1.5m, 2m_foces}. These facilities are also well-suited for repeated or exploratory observations that complement larger programs.

In this work, we present single-epoch high-resolution transmission spectra of the hot Jupiter HD 209458b, observed with the Hanle Echelle Spectrograph (HESP) on the 2.0 m Himalayan Chandra Telescope (HCT). These observations are intended as a feasibility test of HESP’s capability for exoplanet transmission spectroscopy. Given the known challenges for HD 209458b — where previous studies have shown that apparent sodium features can be explained by stellar effects such as the Rossiter–McLaughlin (RM) effect and center-to-limb variation (CLV) (see Section \ref{section3}) — our goal is to assess whether HESP can deliver spectra of sufficient quality to probe such subtle signals. This contributes to evidence supporting the role of moderate telescopes equipped with general-purpose high-resolution spectrographs in exoplanet transmission spectroscopy.

\section{Overview of Telescope and Spectrograph}

The Indian Astronomical Observatory (IAO) is located in Hanle, Ladakh, at a latitude of 32.7797°, a longitude of 78.9639°, and an altitude of approximately 4,500m. The site's low atmospheric pressure, minimal humidity, and very low light pollution make it well-suited for precision astronomical spectroscopy and photometric observations. The 2 m Himalayan Chandra Telescope, located at IAO, is equipped with a fiber-fed Hanle Echelle Spectrograph, the Himalayan Faint Object Spectrograph Camera (HFOSC), and the TIFR Near Infrared Spectrometer and Imager (TIRSPEC)\footnote{\url{https://www.iiap.res.in/centers/iao/facilities/hct/}}. Using HFOSC, both transit photometry and the multi-object low-resolution transmission spectroscopic observations of exoplanets have been successfully conducted at this facility \cite{A_Chakrabarty, suman_hct, athira_hct}.

\begin{table}[h]
    \centering % Center the table on the page
    \begin{tabular}{ll} % Two columns, left-aligned
        \hline
        \multicolumn{2}{c}{Specifications of HCT and HESP Instrument} \\
        \hline
        Primary mirror aperture & 2\,m \\
        Plate scale  & 92.592\,$~\mu\text{m}$ per 1\arcsec{} \\
        Wavelength range & 0.35--1\,$~\mu\text{m}$ \\
        Spectral resolution & 30,000$^\star$ \& 60,000 \\
        Fiber size & 2.7\arcsec{} \\
        \hline
    \end{tabular}
    \caption{This table summarizes key specifications of the Himalayan Chandra Telescope (HCT) and the Hanle Echelle Spectrograph (HESP).
    $^\star$The observing mode used in this work.}
    \label{hcttable}
\end{table}

HESP offers a spectral resolution of R = 60,000 with the image slicer and R = 30,000 without, spanning a broad wavelength range from 350 nm to 1,000 nm in a single exposure (see Table~\ref{hcttable}). In this work, we present an initial demonstration of the use of HESP at HCT for high-resolution transmission spectroscopy, marking an important step toward establishing this facility for exoplanet atmospheric studies.

\label{section2}
\section{Observations}

As part of this project to test the feasibility of HESP, we have planned observations of well-characterized hot Jupiters and ultra-hot Jupiters orbiting bright stars, including HD 209458b, HD 189733Ab, and KELT-9b. In this paper, we focus on HD 209458 b as a benchmark hot Jupiter orbiting a G-type star. Notably, it was the first reported detection of an exoplanet atmosphere \cite{charbonneau1999detection} and remains one of the most extensively studied planets in the literature.

Sodium (Na I) is a key atomic species in exoplanet atmospheric studies due to its strong resonance lines in the optical spectrum and its large scattering cross-section, which makes it highly detectable in transmission spectroscopy. 
These lines probe the upper atmosphere and provide insights into atmospheric composition, thermal structure, dynamics (e.g., winds), and escape processes. Therefore, Na I has become a widely used essential tracer in the characterization and modeling of giant exoplanet atmospheres.

The detection of sodium (Na I) absorption in the transmission spectrum of HD 209458 b has been a subject of considerable debate. Initial low-resolution observations with the Hubble Space Telescope reported the presence of Na I \cite{Charbonneau_2002}, and subsequent studies at both low and high spectral resolutions aimed to confirm its presence using various instruments and telescopes \cite{snellen2004new}. 

However, in 2020, Casasayas-Barris et al. analyzed high-resolution data from the CARMENES and HARPS-N spectrographs and raised significant doubts \cite{na_true}. They suggested that the observed features near the Na I D lines in the transmission spectrum could be attributed not to planetary absorption but to stellar effects—specifically, CLV and RM effect. This non-detection was further supported by two nights of high-resolution, high signal-to-noise observations using ESPRESSO combined with precise stellar modeling \cite{Canocchi2024}.

Although these studies highlight the challenges in conclusively detecting atmospheric sodium due to stellar contamination, there remains significant interest in revisiting this target using different instruments and observing strategies. In this context, our observations aim to assess the performance of the Hanle Echelle Spectrograph (HESP) on the Himalayan Chandra Telescope (HCT) for ground-based high-resolution transmission spectroscopy.

In this study, we conducted observations of HD 209458 b using HESP at a spectral resolution of 30,000. Each exposure lasted 20 minutes and achieved a signal-to-noise ratio of about 65 in the continuum near the Na I D lines. A total of 15 frames were obtained, spanning both in-transit and out-of-transit phases. These observations were carried out in October 2023 under proposal code HCT-2023-C3-P38.

\begin{table}[h!]
\centering
\caption{Stellar and planetary parameters of HD~209458 b used in this work}
\label{tab:rmclv_params}
\begin{tabular}{llcl}
\hline
\textbf{Parameter} & \textbf{Description} & \textbf{Value} & \textbf{Reference} \\
\hline
\multicolumn{4}{c}{\textit{Stellar Parameters}} \\
$T_{\mathrm{eff}}$ & Effective temperature & 6065 ± 50 K & Torres et al. (2008)\cite{torres2008accurate} \\
$\log g$ & Surface gravity (cgs) & 4.361 ± 0.007 & Torres et al. (2008)\cite{torres2008accurate} \\
$\mathrm{[Fe/H]}$ & Metallicity & 0.00 ± 0.05 & Torres et al. (2008)\cite{torres2008accurate} \\
$R_\star$ & Stellar radius & 1.155$^{+0.014}_{-0.016}$ $R_\odot$ & Torres et al. (2008)\cite{torres2008accurate} \\
$v \sin i_\star$ & Projected rotation velocity & 4.70 ± 0.16 km\,s$^{-1}$ & Winn et al. (2005)\cite{winn2005spinorbit} \\

\hline
\multicolumn{4}{c}{\textit{Planetary Parameters}} \\
$P$ & Orbital period & 3.52474859 ± 0.00000038 days & Bonomo et al. (2017)\cite{bonomo2017}\\
$T_0$ & Transit mid-time (BJD) & 2454560.80588 ± 0.00008 & Evans et al. (2015)\cite{evans2015} \\
$a/R_\star$ & Scaled semi-major axis & 8.76 ± 0.04 & Torres et al. (2008)\cite{torres2008accurate} \\
$i$ & Orbital inclination & 86.71 ± 0.05$^\circ$ & Torres et al. (2008)\cite{torres2008accurate} \\
$\lambda$ & Sky-projected spin-orbit angle & $-1.6 \pm 0.3^\circ$ & Casasayas-Barris et al. (2020)\cite{na_true} \\
$R_p/R_\star$ & Planet-to-star radius ratio & 0.1240 $\pm$ 0.00046 & Beaulieu et al. (2010)\cite{Beaulieu_2010} \\
$b$ & Impact parameter & $0.5000 \pm 0.007$ & Brandeker et al.(2022)\cite{Brandeker2022} \\
$K_\mathrm{planet}$ & Radial velocity semi-amplitude & 144.89 km\,s$^{-1}$ & Casasayas-Barris et al. (2020)\cite{na_true} \\
\hline
\end{tabular}
\end{table}

\label{section3}
\section{Data Reduction \& Analysis}

\label{section4}

After obtaining the two-dimensional echelle spectra, we performed standard preprocessing steps, such as bias correction and cosmic ray removal, using the HESP pipeline\footnote{\url{https://github.com/arunsurya77/hesp_pipelinee}}. We did not apply flat-field correction because our analysis is differential, as the in-transit spectra are divided by the median-combined out-of-transit spectra (see Section \ref{section7}).

For spectral extraction and wavelength calibration (using Th-Ar lamps), we employed the \texttt{apall}, \texttt{ecidentify}, and \texttt{ecdispcor} tasks from the \texttt{echelle} package in PyRAF\footnote{\url{https:/pyraf.readthedocs.io/en/latest/}}. We achieved a root mean square (RMS) residual of 0.0017 using a 5th-order Chebyshev fit in both the x and y directions for the wavelength calibration, based on approximately 350 reference features distributed across 45 echelle orders.

The Na I D lines fall within two echelle orders; however, for our analysis, we considered only the order free from artifacts in the sodium region. Although cosmic ray removal was performed by the HESP pipeline, a few emission features remained in the selected order. We removed these residual features manually using the \texttt{splot} task in PyRAF.

\subsection{Telluric Correction}
\label{section5}

High-resolution, ground-based transmission spectroscopy faces significant challenges when probing exoplanetary atmospheres, primarily due to contamination from telluric lines. These spectral features, typically observed as absorption, originate in Earth’s atmosphere as starlight interacts with atmospheric molecules during its passage through the atmosphere.

\begin{figure}[h]
    \centering
    \includegraphics[width=0.6\textwidth]{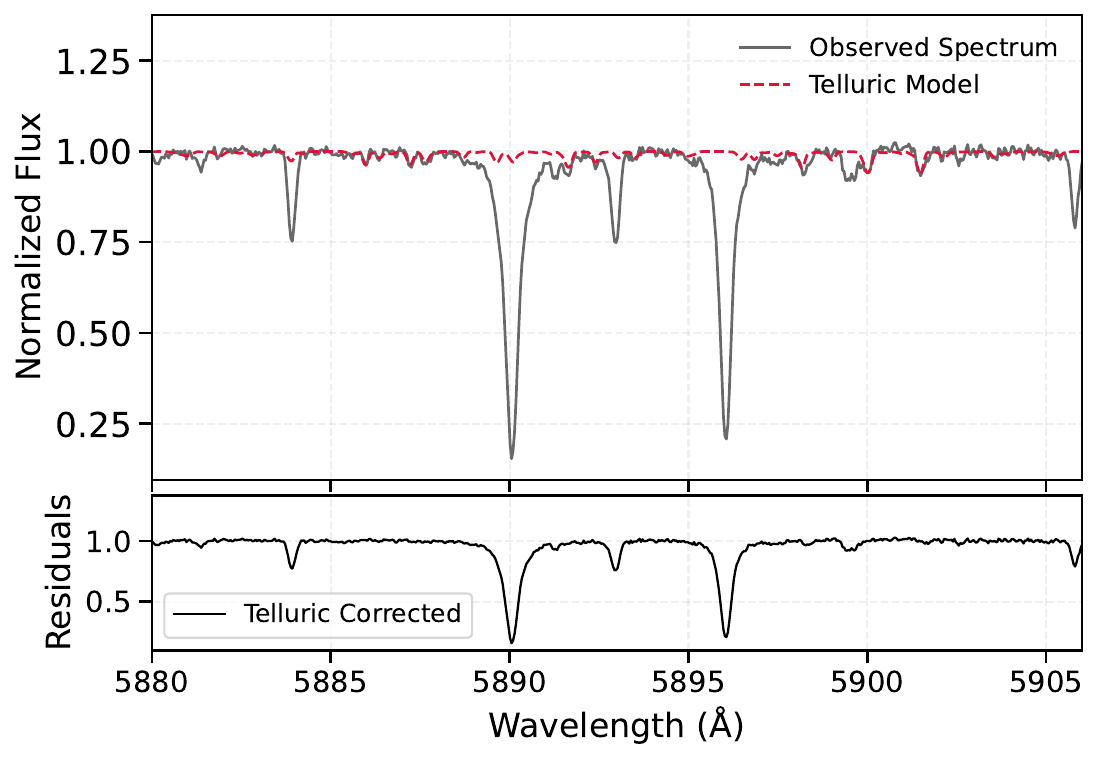}
    \caption{Top panel: One of the observed spectra in the observatory rest frame (black) along with the telluric model generated by \texttt{telFit} (red). Bottom panel: The observed spectrum after correcting for telluric absorption lines.}
    \label{telluric_plot}
\end{figure}

Removing telluric contamination is one of the most important steps in high-resolution transmission spectroscopy. To address this, we employed the Python package \texttt{telfit}\footnote{\url{https://github.com/kgullikson88/Telluric-Fitter}}\cite{telfit} to model telluric lines in the observatory rest frame using wavelength-calibrated observed spectra. The observed spectra provided to \texttt{telfit} were normalized using a continuum obtained by fitting the spectrum with a default seventh-order polynomial, and the parameters were chosen based on typical seasonal conditions at IAO: pressure ranging from 570 to 590 hPa, temperature between 270 and 290 K, relative humidity from 0 to 60\%, and the zenith angle for each exposure was computed at the midpoint of the exposure time using the Python package \texttt{Astroplan}\footnote{\url{https://github.com/astropy/astroplan}}.

Telluric contamination was removed by dividing the observed spectra by the modeled telluric spectrum\footnote{\url{https://github.com/bestha95/telfit_telluric_correction}}. The telluric model and the corrected spectra are shown in Figure \ref {telluric_plot}.

\subsection{Velocity Corrections}

\begin{figure}[h]
    \centering
    \subfigure[]{
        \includegraphics[width=0.45\textwidth]{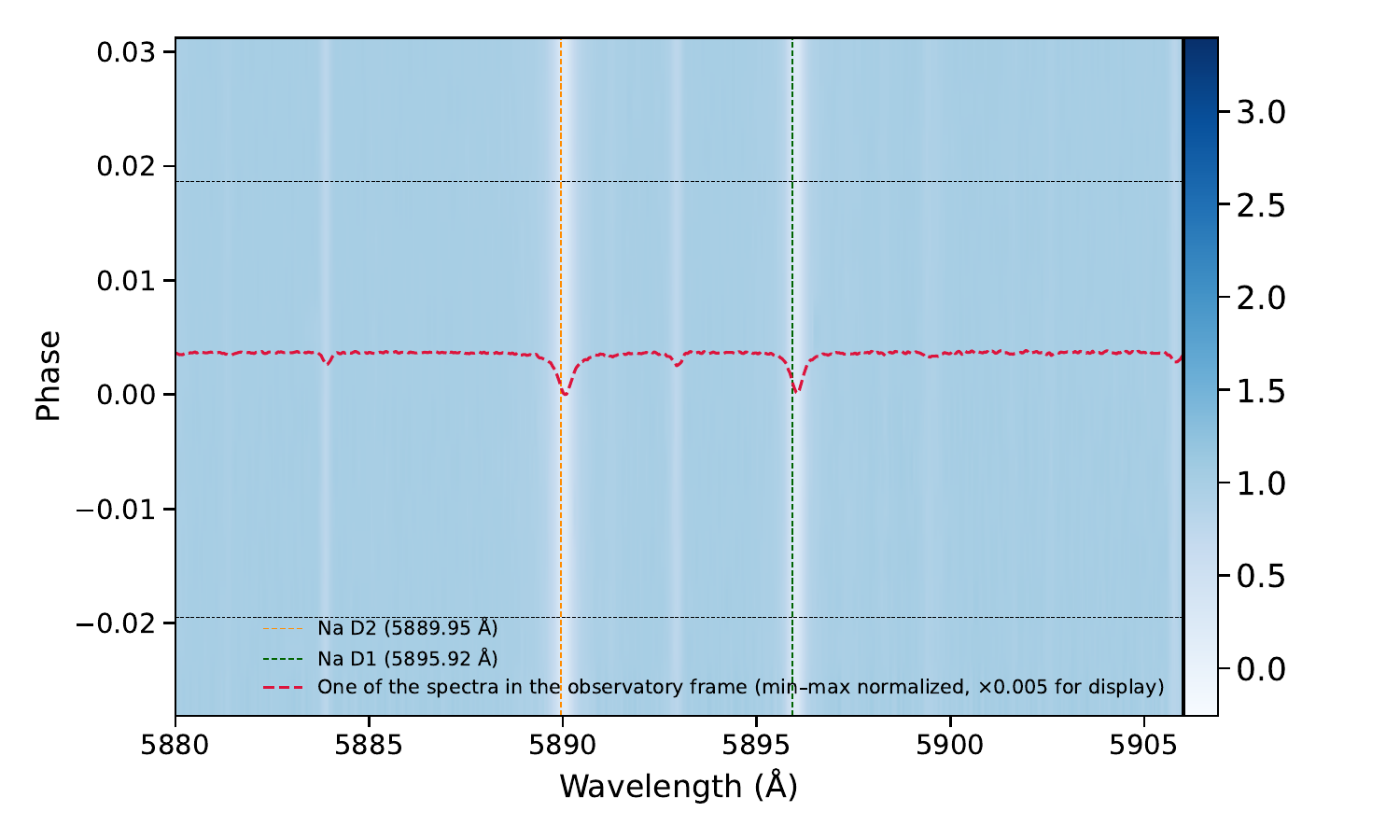}
    }
    \hfill
    % Second image
    \subfigure[]{
        \includegraphics[width=0.45\textwidth]{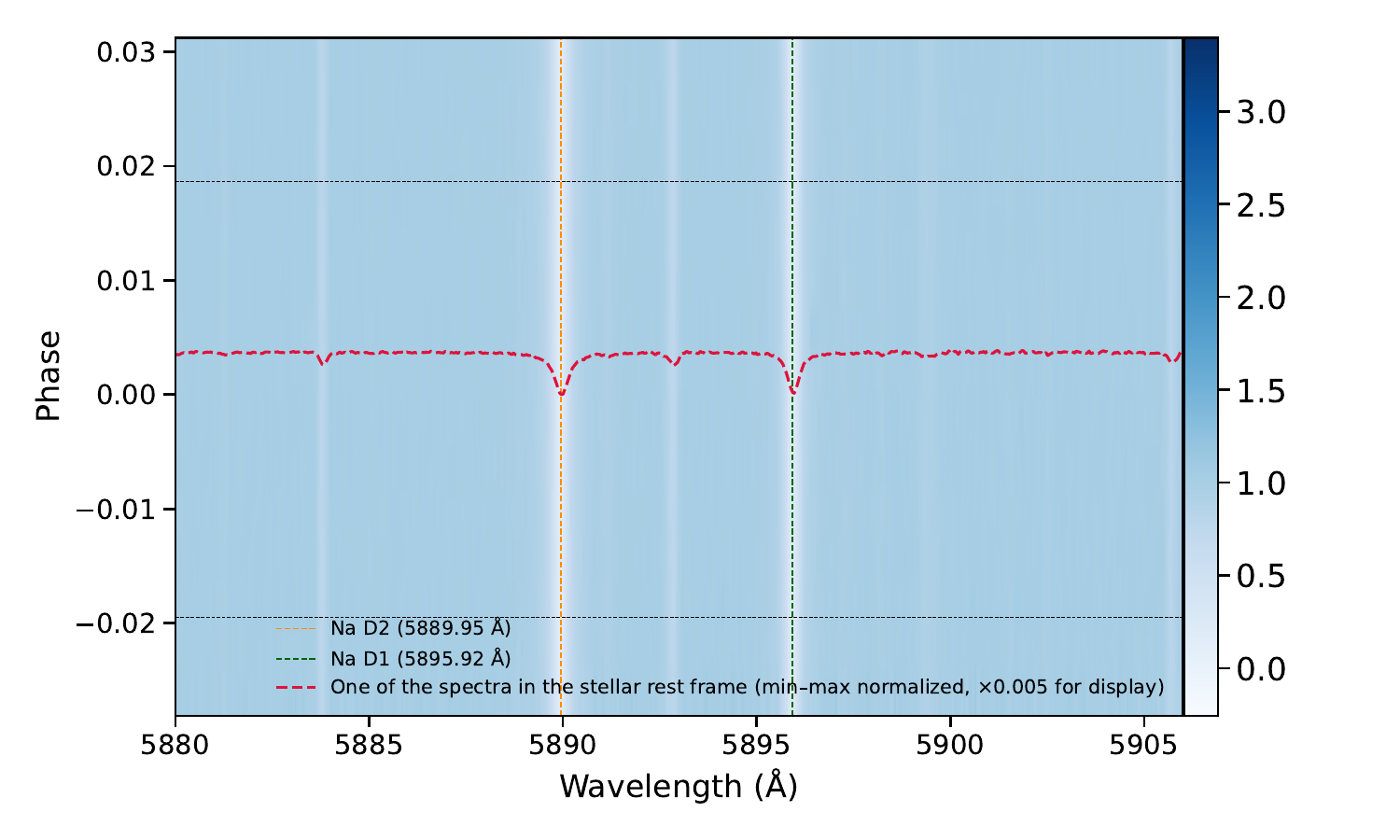}
    }

    \caption{Comparison of sodium line profiles in the observatory frame and the stellar rest frame. The vertical orange and green dotted lines mark the positions of the Na D2 and D1 lines, respectively. The region between the two horizontal dotted lines indicates in-transit observations, while the regions outside represent out-of-transit data in both panels (a) and (b). Panel (a) displays a 2D image of the Na D1 and D2 regions across orbital phase in the observatory frame. Panel (b) shows the same data after being shifted to the stellar rest frame. A spectrum is overplotted as a red dotted line in the middle of panel (a \& b) to verify the alignment of the sodium line positions.}
    \label{orsr}
\end{figure}

The telluric-corrected spectra are initially in the observatory rest frame. To shift them into the stellar rest frame, we apply barycentric velocity corrections (calculated using the \texttt{pyAstronomy} package), along with the systemic velocity and stellar orbital velocity corrections, as described by Equation \ref{srf_eq}. The calculated total stellar rest frame velocities across the exposures range from approximately 4~km~s$^{-1}$ to 5~km~s$^{-1}$. To align all spectra in the stellar rest frame, we interpolated flux values onto a velocity-corrected wavelength grid (see Figure \ref{orsr}).

The correction to the stellar rest frame is given by:
\begin{equation}
\lambda_{\mathrm{rest,\,\star}} = \lambda_{\mathrm{obs}} \left(1 - \frac{v_{\mathrm{SRF}}}{c} \right)
\label{srf_eq}
\end{equation}

The total radial velocity relative to the observer is:
\begin{equation}
v_{\mathrm{SRF}} = -v_{\mathrm{BERV}} + v_{\mathrm{sys}} + v_{\star}(t)
\end{equation}

The stellar orbital velocity due to the planet is:
\begin{equation}
v_{\star}(t) = K_{\star} \sin\left(2\pi \frac{t - T_0}{P} \right)
\end{equation}

\noindent
\textbf{where:}
\begin{align*}
\lambda_{\mathrm{rest,\,\star}} &\quad \text{: Wavelength in the stellar rest frame} \\
\lambda_{\mathrm{obs}}          &\quad \text{: Observed wavelength} \\
v_{\mathrm{SRF}}                &\quad \text{: Total stellar radial velocity relative to observer} \\
v_{\mathrm{BERV}}               &\quad \text{: Barycentric Earth radial velocity} \\
v_{\mathrm{sys}}                &\quad \text{: Systemic velocity of the star} \\
v_{\star}(t)                    &\quad \text{: Star's orbital velocity due to planet at time } t \\
K_{\star}                       &\quad \text{: Stellar radial velocity semi-amplitude} \\
T_0                             &\quad \text{: Time of mid-transit (inferior conjunction)} \\
P                               &\quad \text{: Orbital period of the planet} \\
t                               &\quad \text{: Time of observation} \\
c                               &\quad \text{: Speed of light}
\end{align*}

\subsection{RM and CLV Modeling}
\label{section6}

The Rossiter–McLaughlin (RM) effect is a spectroscopic anomaly observed during a planetary transit. As the planet moves across the rotating stellar disk, it sequentially blocks regions with different projected rotational velocities. This results in a temporary distortion of the observed stellar radial velocity: an apparent redshift occurs when the planet obscures the blueshifted limb, followed by an apparent blueshift as it crosses the redshifted limb.

At the same time, the planet also blocks light from regions of the stellar disk that have different viewing angles, leading to variations in the observed stellar spectral line shapes and intensities. The central regions of the stellar disk contribute stronger and narrower lines, while the limb regions produce weaker and broader lines. This phenomenon is known as center-to-limb variation (CLV).

Together, RM and CLV effects distort the observed stellar lines in high-resolution transmission spectroscopy. Accurately modeling these effects is essential to isolate and interpret the true exoplanet atmospheric signal.

To model the combined RM and CLV effects, we employed a forward-modeling approach using the \texttt{PyAstronomy} package. We retrieved specific intensity spectra across the stellar disk via the \texttt{spectralLib} interface \cite{pyastronomySpecLib}. In our initial modeling, we adopted a stellar model with parameters $T_{\rm eff} = 6250$~K, $\log g = 4.5$, and [Fe/H] = 0.0, sourced from the \texttt{spectralLib} database. However, these values are inconsistent with the well-established parameters for HD~209458 ($T_{\rm eff} = 6065 \pm 50$~K, $\log g = 4.361 \pm 0.007$, see Table~2). Since the center-to-limb variation (CLV) is highly sensitive to effective temperature and surface gravity, the RM+CLV model presented here should be considered illustrative of the approximate magnitude and structure of the effect, rather than a precise quantitative prediction for this system. A definitive analysis requires a model using the correct stellar parameters.

The modeled spectra wavelength-filtered around the Na~D doublet region and normalized using third-order polynomial fitting via the \texttt{spectrum\_overload}\footnote{\url{https://pypi.org/project/spectrum-overload/}} package and continuum estimation\footnote{\url{https://github.com/bestha95/RM_CLV}}.

The RM+CLV model was generated at the native high resolution of the \texttt{spectralLib} grid. To match the instrumental resolution of HESP ($R \approx 30{,}000$), we degrade this model by convolving it with a Gaussian kernel. The full width at half maximum (FWHM) of the kernel is set to 3.2 pixels, as measured from emission lines in the Th-Ar lamp spectra near the D1 and D2 line regions taken before the observations.

\begin{figure}[h!]
    \centering
    \includegraphics[width=0.8\textwidth]{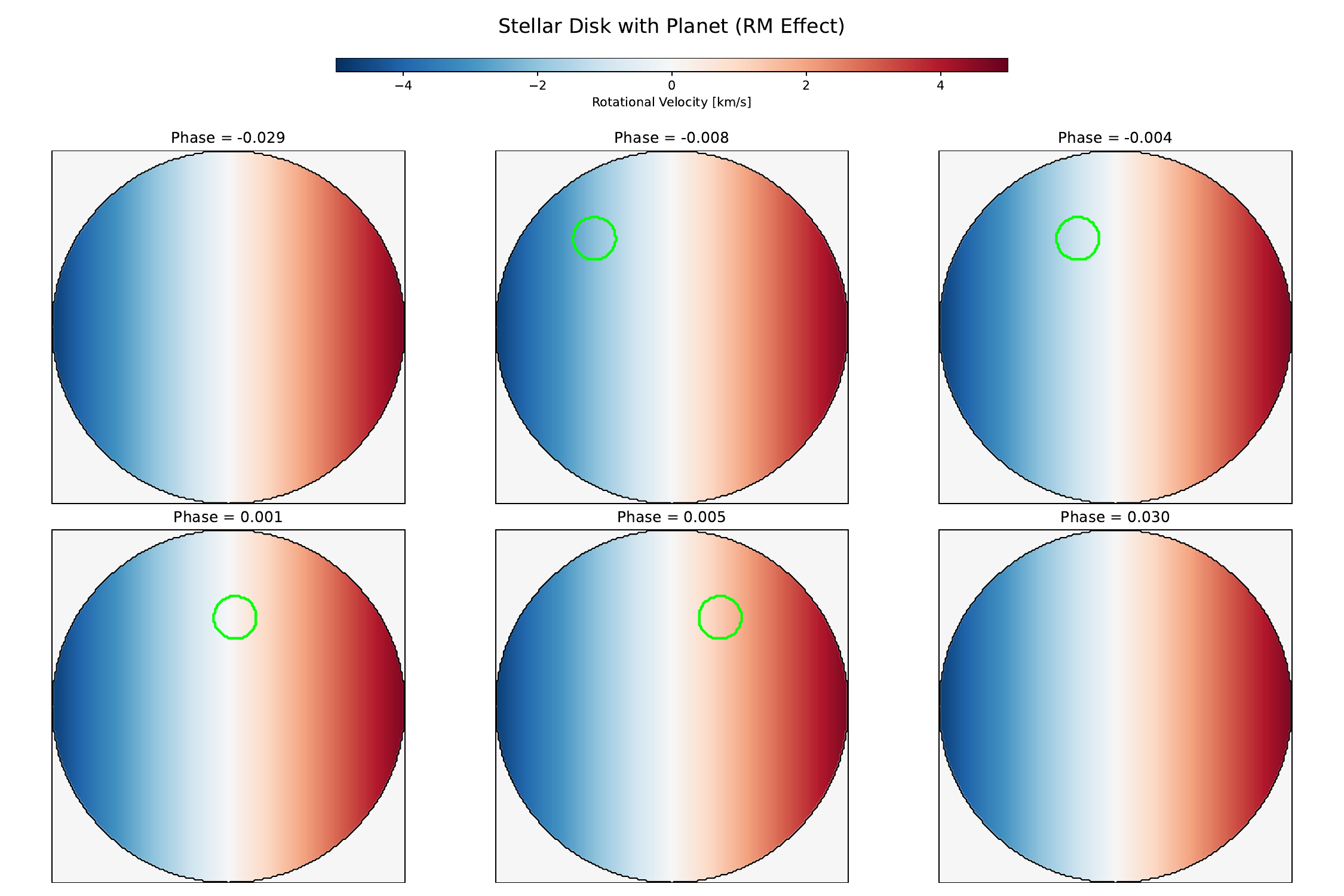}
    \caption{Simulated views of the stellar disk with rotational velocity indicated by color: blue for approaching (blueshifted) regions and red for receding (redshifted) regions due to stellar rotation. The green circle represents the transiting planet. The panels show the simulated position of the planet on the stellar disk at different orbital phases. Only a subset of the full phase coverage is shown here for illustration.}
    \label{fig:rm_disk_views}
\end{figure}

\begin{figure}[h!]
    \centering
    \includegraphics[width=0.6\textwidth]{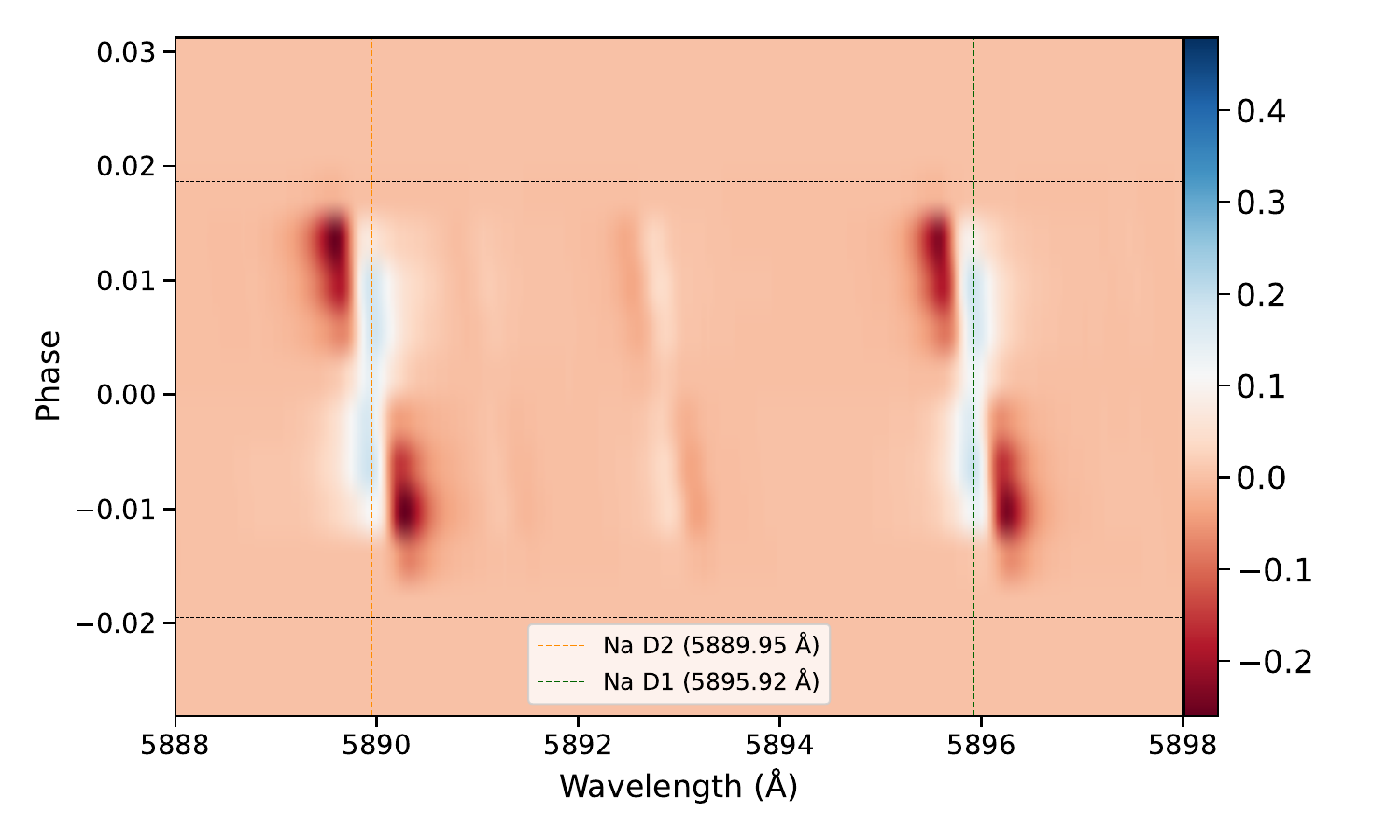}
    \caption{RM+CLV modeled spectra after dividing by the out-of-transit spectrum and shifted to the planetary rest frame. Vertical orange and green dotted lines indicate the positions of the Na D2 and D1 lines, respectively. The region between the two horizontal dotted lines corresponds to in-transit, while regions outside represent out-of-transit.}
    \label{fig:rm}
\end{figure}

\begin{figure}[h!]
    \centering
    \includegraphics[width=1\textwidth]{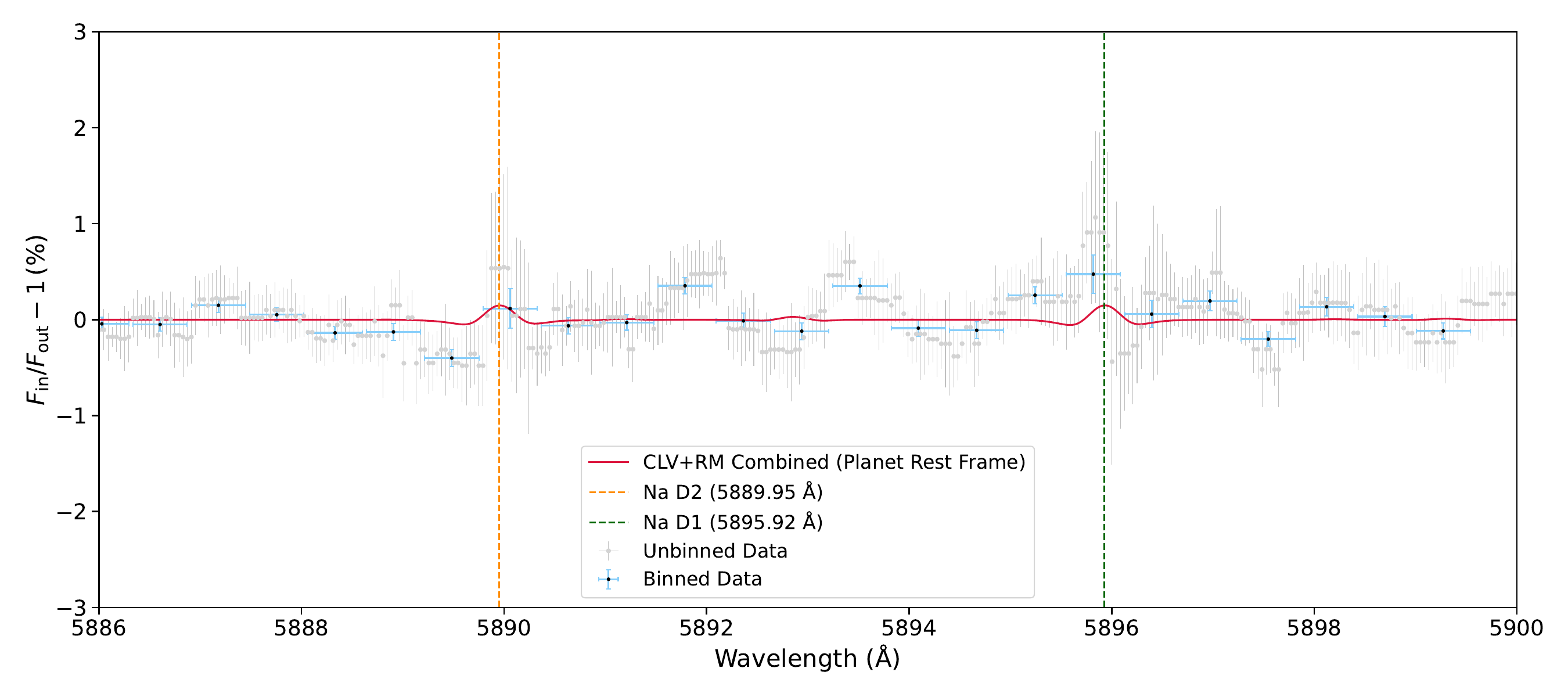}
    \caption{The blue points represent the binned data, with each bin containing 14 points (corresponding to a bin width of 0.575\,\AA). Vertical error bars denote the standard error of the mean calculated within each bin, while the horizontal error bars represent the bin width. The red line shows the mean RM+CLV model. The orange and green dotted lines indicate the positions of the Na D2 and D1 lines, respectively.}
    \label{fig:transmission}
\end{figure}

A two-dimensional stellar disk was constructed using a Cartesian grid with a resolution of 0.01 stellar radii per pixel, covering the full stellar disk. Each pixel was assigned a local $\mu$ value and Doppler shifted based on the projected rotational velocity. For each time stamp, the planet's position was calculated using Keplerian orbital parameters, and the corresponding occulted pixels on the stellar disk were masked (see Figure \ref{fig:rm_disk_views}). The in-transit spectrum was synthesized by integrating the Doppler-shifted and limb-dependent intensities from the unocculted portion of the disk. These spectra were normalized using the median out-of-transit spectrum and then shifted into the planet’s rest frame using Equation~\ref{prf} (see Figure~\ref{fig:rm}). The average of the in-transit spectra, excluding ingress and egress phases, was used as the RM+CLV model (see Figure~\ref{fig:transmission}).

\section{Transmission Spectrum}
\label{section7}

\begin{figure}[h!]
    \centering
    \includegraphics[width=0.6\textwidth]{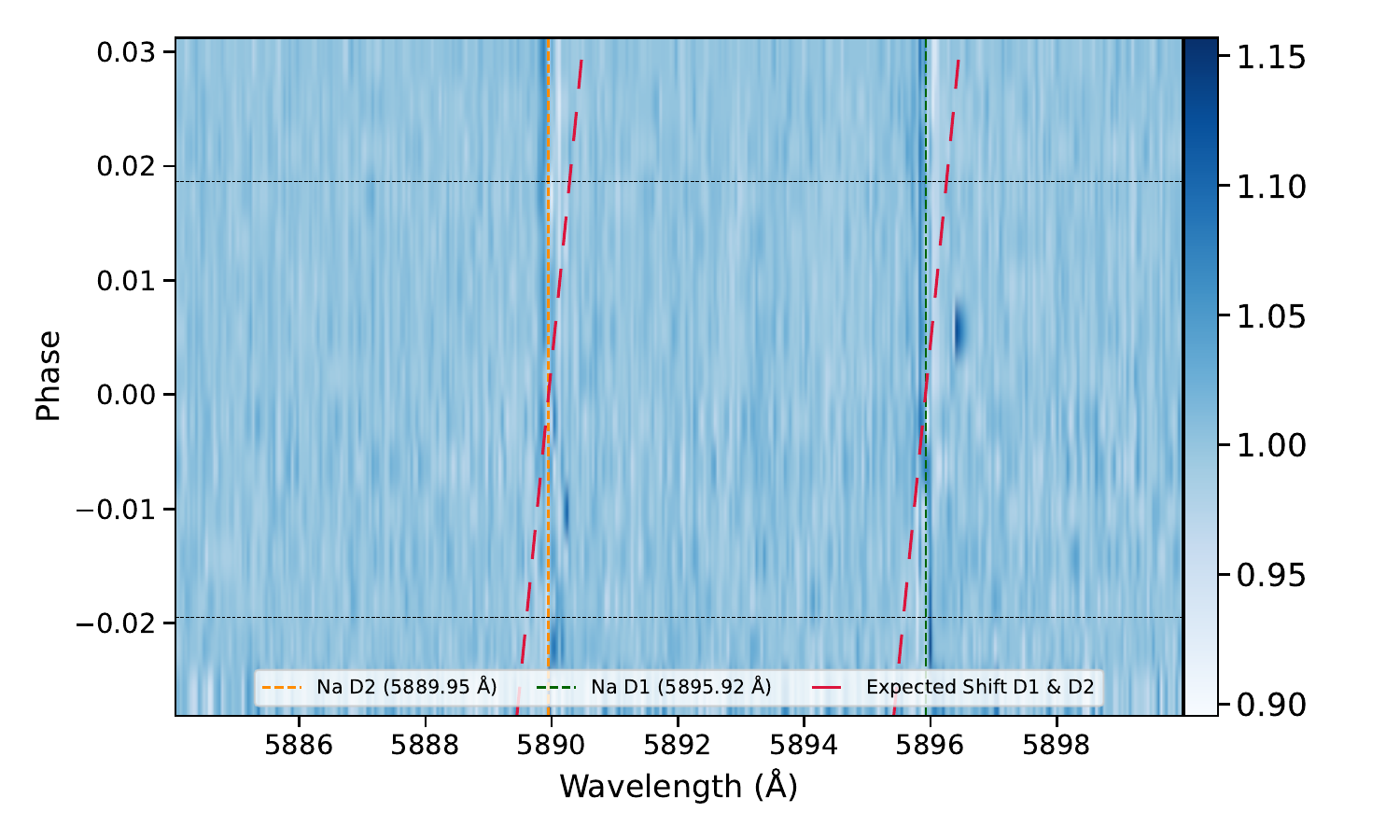}
    \caption{Observed spectra after dividing by the out-of-transit spectrum (residuals), shown in the stellar rest frame. Vertical orange and green dotted lines indicate the positions of the Na D2 and D1 lines, respectively. The region between the two horizontal dotted lines corresponds to in-transit observations, while regions outside represent out-of-transit data. The tilted red lines at the positions of Na D2 and D1 indicate the expected shift of the planetary absorption signal.}
    \label{fig:na_residual_stellar_frame}
\end{figure}

We combined the out-of-transit spectra to construct a master out-of-transit spectrum. Each individual spectrum was then divided by this master spectrum to remove the stellar contribution. The resulting residuals were Doppler-shifted into the planetary rest frame using Equation \ref{prf} to account for the planet's orbital motion \cite{athira_hrs}, as shown in Figure \ref{fig:na_residual_stellar_frame}. Residuals corresponding to ingress and egress phases, as well as out-of-transit data, were excluded, and the remaining in-transit residuals were weighted-averaged using the inverse variance as weights to enhance the signal-to-noise ratio\cite{Langeveld}.

Doppler shift to planetary rest frame:

\begin{equation}
\lambda_{\mathrm{rest,\,p}} = \lambda \left(1 - \frac{v_{\mathrm{p}}(t)}{c} \right)
\label{prf}
\end{equation}
\begin{equation}
v_{\mathrm{p}}(t) = K_{\mathrm{p}} \sin\left(2\pi \frac{t - T_0}{P} \right)
\end{equation}

\noindent
\textbf{where:}
\begin{align*}
\lambda_{\mathrm{rest,\,p}} &\quad \text{: Wavelength in the planet's rest frame} \\
\lambda                     &\quad \text{: Wavelength in the stellar rest frame} \\
v_{\mathrm{p}}(t)           &\quad \text{: Planet's radial velocity at time } t \\
K_{\mathrm{p}}              &\quad \text{: Planet's radial velocity semi-amplitude} \\
T_0                         &\quad \text{: Time of mid-transit (inferior conjunction)} \\
P                           &\quad \text{: Orbital period of the planet}
\end{align*}

To remove broad variations, a 10-point median filter was applied to the residual spectrum. To further enhance the visibility of spectral features and suppress noise, the filtered residuals were binned into fourteen discrete wavelength points with a bin width of $0.575\,\text{\AA}$.

\section{Conclusion and Future Work}

In this work, we presented the attempt to use HESP for high-resolution transmission spectroscopy of HD~209458\,b. Our initial analysis revealed subtle, apparent residuals of approximately 0.1\% and 0.5\% at the locations of the Na D2 and D1 lines, respectively (see Figure~\ref{fig:transmission}). For context, we compared these values to a preliminary model of the expected stellar contamination (RM and CLV effects), which predicts features of a similar order of magnitude. Due to the limited exposures from this single-epoch observation, we did not attempt to construct a transmission light curve. Future observations spanning multiple epochs would be essential for light curve construction.

\newpage
Our analysis also shows noticeable variation and wiggles in the weighted-averaged spectrum, likely caused by systematics. In future work, applying techniques such as \texttt{SYSREM}\cite{pyastronomySysrem}—which has been used in high-resolution transmission spectroscopy—or two-dimensional Gaussian process modeling\footnote{\url{https://github.com/markfortune/luas}}—an approach that has been recently applied in low-resolution transmission spectroscopy—could help mitigate these variations. 

While HESP is not specifically optimized for high-precision exoplanet spectroscopy, this preliminary result demonstrates its potential to contribute to atmospheric studies, particularly when combined with careful analysis and multi-epoch observations. Extending this approach to additional Hot Jupiters using HESP–HCT would further support the feasibility of conducting high-resolution exoplanet atmosphere characterization from the Indian Astronomical Observatory.

%\conclusion
\acknowledgments % equivalent to \section*{ACKNOWLEDGMENTS}  

We thank the staff of the Indian Astronomical Observatory (IAO), Hanle, and the Center for Research \& Education in Science \& Technology (CREST), Hoskote, for their support in making these observations possible. The facilities at IAO and CREST are operated by the Indian Institute of Astrophysics (IIA), Bangalore. We especially acknowledge the assistance of Ms. Yogita Patel from the Hoskote campus for her valuable support during the observations. We also thank the HCT Time Allocation Committee for granting observing time for this project.

% References
\bibliography{report} %bibliography data in report.bib

\bibliographystyle{spiebib} % makes bibtex use spiebib.bst
\end{document}